\documentclass[12pt]{iopart}

\begin{document}

\title[Non--Localizability of Electric Coupling]{Non--Localizability of Electric Coupling and
 Gravitational Binding of Charged Objects}

\author{Matthew Corne $^{1}$, Arkady Kheyfets $^{1}$ and Warner A. Miller $^{2}$}

\address{$^{1}$ Department of Mathematics, North Carolina State University, Raleigh, NC 27695-8205}
\address{$^{2}$ Department of Physics, Florida Atlantic University, Boca Raton, FL 33431-0991}

\eads{\mailto{macorne@ncsu.edu}, \mailto{kheyfets@ncsu.edu}, \mailto{wam@physics.fau.edu}}

\begin{abstract} 
The energy--momentum tensor in general relativity contains 
only localized contributions to the total energy--momentum. Here, we consider a static, 
spherically symmetric object consisting of a charged perfect fluid. For this object, 
the total gravitational mass contains a non--localizable contribution of electric coupling 
(ordinarily associated with electromagnetic mass). We derive an explicit expression for 
the total mass which implies that the non--localizable contribution of electric coupling 
is not bound together by gravity, thus ruling out existence of the objects with pure 
Lorentz electromagnetic mass in general relativity.

\bigskip

\end{abstract}

\bigskip

In general relativity, the energy--momentum tensor is determined as 
the variational derivative of the matter lagrangian with respect to the spacetime metric. 
For systems that include charged matter (either charged particles or charged fluid), 
the matter lagrangian is composed of the lagrangian of particles, lagrangian of the 
electromagnetic field, and the interaction lagrangian (interaction between the particles 
or fluid and the electromagnetic field). An interesting feature of the variational 
procedure that yields the energy--momentum tensor is that the interaction term does not 
contribute to the energy-momentum. The total energy--momentum tensor is the sum of the 
energy--momentum of the particles (or fluid) and the energy--momentum of the electromagnetic 
field. This does not mean that the electromagnetic coupling between particles (or inside 
the fluid) does not contribute to the total energy of the system. The correct 
interpretation of the situation is that the contribution of the electromagnetic coupling 
is not localizable; it cannot be described by the energy--momentum tensor, which represents 
the distribution density (hence localization) of energy--momentum--stress. Rather, it should 
result from integration of some quantity over the object. 

This feature is by no means unique for electromagnetic coupling. For instance, it takes place in
gravitational coupling. The gravity field itself does not contribute to the energy--momentum tensor
but, in the theory of neutral spherical stars, does provide a contribution to the total gravitational
mass of the star which becomes obvious only in the process of integration of the 00--component of
Einstein's equations over the star. 

To see how this issue is resolved for electromagnetic coupling, we consider the simplest
generalization of the spherically symmetric, static neutral star which can be called the
charged star. Objects of this kind have been investigated previously \cite{Ndu,KB,Sin}. The principal 
domains of application are the search for charged astrophysical objects \cite{PS,RD} and the 
general relativistic generalization of the Lorentz theory of the extended electron model \cite{TRK,Gau,Bon}, 
the second of which was based originally on a belief that, in general relativity, gravitational 
binding might stabilize a pure electromagnetic object. A very simple, model independent analysis 
below shows that it is impossible in principle.  The resulting expression for the gravitational mass of a 
charged star is not meant to help in generating new solutions and does not do that as it can be evaluated 
only after a particular model has been solved. Instead, it provides a physically transparent description 
of the total gravitational mass composition and allows for drawing important conclusions common for all models. 
    
We assume here the star to be formed by a charged perfect fluid. The energy-momentum
tensor for this fluid is given by \cite{MTW} 

\begin{equation}
T_{\mu\nu} = (\mu + p)u_{\mu}u_{\nu} + g_{\mu\nu} p + 
\frac{1}{4\pi}\left({F_{\mu}}^{\kappa}F_{\nu\kappa} - 
\frac{1}{4}F_{\kappa\lambda}F^{\kappa\lambda} g_{\mu\nu}\right) 
\end{equation}
where \(\mu\) and \( p\) are the (proper) density of the fluid and the pressure of the fluid.
The notation \(\rho\) is reserved for the (proper) density of the charge. 

For a spherically symmetric, static gravitational field there is a coordinate system in which
the spacetime line element is expressed by 

\begin{eqnarray}
ds^{2} = -e^{2\Phi(r)}dt^{2} + e^{2\Lambda(r)}dr^{2} + 
r^{2}(d\theta^{2}+\sin^{2}\theta d\phi^{2}) \nonumber\\
\Rightarrow g_{00} = -e^{2\Phi(r)}; \: g_{rr} = e^{2\Lambda(r)}; \: 
g_{\theta\theta} = r^{2}; \: g_{\phi\phi} = r^{2}\sin^{2}\theta
\end{eqnarray}
with $ g_{00}g_{rr} = -1 $ \cite{TRK}. 

In what follows, we are interested only in the 00--component of Einstein's equations. The
standard procedure for computing the Einstein tensor (done here in the coordinate frame)
yields the expression for \( G_{00}\) \cite{Schutz}

\begin{equation}
G_{00} = \frac{1}{r^{2}}e^{2\Phi(r)}\frac{d}{dr}[r(1-e^{-2\Lambda(r)})].
\end{equation} 

The same symmetries, when applied to the matter forming the star, are reduced to the
assumptions that \(\mu = \mu (r)\), \(\rho = \rho (r)\), \(p = p(r)\), and that the 
electromagnetic field is a radial electric field \(\mathbf{E}\) that can be written 
as \(\mathbf{E} = E(r) \mathbf{e}_{\hat{r}}\) where \(\mathbf{e}_{\hat{r}}\) is the 
radial vector of the physical (normalized) frame associated with the coordinate frame. 
The \( T_{00}\) component of the energy--momentum tensor is given by 

\begin{equation} 
T_{00} = e^{2\Phi(r)}\left[\mu(r) + \frac{\mathbf{E}^{2}}{8\pi}\right].
\end{equation}

The relevant equation, $ G_{00} = 8\pi T_{00} $, is expressed by 

\begin{equation}
\frac{d}{dr}\left[r(1 - e^{-2\Lambda(r)})\right] = 8\pi r^2 \mu(r) + r^2 \mathbf{E}^{2}.
\end{equation} 

The electric field is determined by the charge distribution via Maxwell's equations. 
In view of the symmetries of the problem, all but one of Maxwell's equations are satisfied 
trivially. The remaining equation is

\begin{equation}
\label{divE} 
\nabla\cdot\mathbf{E} = 4 \pi \rho .
\end{equation} 

This equation, together with the divergence theorem applied to the infinitesimal spherical 
layer between \( r\) and \( r + dr\), yields the expression (see Appendix A)

\begin{equation}\label{EField} 
E(r)  = \frac{q(r)}{r^2} 
\end{equation} 
where \( q(r)\) is the charge inside the sphere of radius \( r\),

\begin{equation} 
q(r) = \int\limits_0^r \rho(r) 4\pi r^2 {\rm e}^{\Lambda (r)} dr .
\end{equation} 

Then, (5) can be rewritten as 

\begin{equation} 
\frac{d}{dr}\left[ r(1 - e^{-2\Lambda (r)})\right] = 8\pi r^2 \mu (r) + 
\frac{q^2(r)}{r^2} .
\end{equation} 

The subsequent analysis of this equation continues along the same lines as that of a neutral 
star and involves replacing the function \( \Lambda (r)\) by a new function \(\tilde m(r)\) 

\begin{equation} 
\tilde m(r) = \frac{1}{2} r \left( 1 - e^{-2\Lambda(r)}\right) 
\end{equation} 
which reduces (5) to 

\begin{equation} 
\frac{d\tilde m}{dr} = 4\pi r^2 \mu (r) + 
\frac{q^2(r)}{2 r^2} . 
\end{equation} 

The boundary of the star \( r = R\) is determined by the conditions \( \mu (r) = \rho (r) = 0\) 
for \( r \geq R\) which imply, in particular, that outside of the star \(q (r)\) is a constant 

\begin{equation} 
q(r) = q(R) = Q = \int\limits_0^R \rho(r) 4\pi r^2 {\rm e}^{\Lambda (r)} dr 
\end{equation} 
that can be identified as the total charge of the star. 

The relevant Einstein equation outside the star  

\begin{equation} 
\frac{d\tilde m}{dr} = \frac{Q^2}{2 r^2} 
\end{equation}
is easy to integrate 

\begin{equation} 
\tilde m(r) = -\frac{Q^2}{2 r} + M 
\end{equation}
where \( M\) is the constant of integration that has physical interpretation as the total 
gravitational mass of the star, since substitution of 

\begin{equation}
e^{2\Lambda(r)} = \frac{1}{1 - \frac{2\tilde m(r)}{r}} = 
\frac{1}{1 - \frac{2M}{r} + \frac{Q^{2}}{r^{2}}} 
\end{equation} 

\noindent in the line element turns it into the Reissner--Nordstr{\o}m metric, which identifies 
\( M\) as the gravitational mass of the star based on the analysis of the Keplerian motion of 
neutral test particles around the star. 

The composition of the gravitational mass can be revealed by extending the similar analysis of 
the Einstein equation to the interior of the star. We introduce a new function \( m(r)\) such that 

\begin{equation} 
\tilde m(r) = -\frac{q^2(r)}{2 r} + m(r) 
\end{equation} 

\noindent which reduces the Einstein equation to 

\begin{equation} 
\frac{d}{dr}\left( m(r) - \frac{q^2(r)}{2 r}\right) = 4 \pi r^2 \mu (r) + \frac{q^2(r)}{2 r^2}.
\end{equation} 

Outside the star (where \(\mu (r)= 0\) and \( q (r) = Q = \textrm{Const.}\)), this equation remains in full 
agreement with our previous considerations and implies that \( m(r) = m(R) = M = \textrm{Const.}\) 
for \( r \geq R\). However, inside the star, \(\mu (r) \neq 0\) and \( q(r)\) are not constant,

\begin{equation} 
\frac{dq}{dr} = \rho (r)\,  4 \pi r^2 {\rm e}^{\Lambda (r)}.
\end{equation} 

\noindent Then, the Einstein equation takes the form 

\begin{equation} 
\frac{dm}{dr} = 4\pi r^2 \mu (r) + \frac{q(r)\, \rho (r)\,  4\pi r^2 {\rm e}^{\Lambda (r)}}{r} .
\end{equation} 

Integration of this equation yields \( m(r)\), which is called sometimes the mass function, 

\begin{equation} 
m(r) = \int\limits_0^r 4\pi r^2 \mu (r) dr + \int\limits_0^r \frac{q(r)\, \rho (r)\, 
4 \pi r^2 {\rm e}^{\Lambda (r)}}{r}\, dr 
\end{equation}

\noindent although it cannot be interpreted as the mass--energy inside \( r\) since total 
energy is not localizable in general relativity. 

However, the value  

\begin{equation} 
M = m(R) = \int\limits_0^R 4\pi r^2 \mu (r) dr + 
\int\limits_0^R \frac{q(r)\, \rho (r)\,  
4 \pi r^2 {\rm e}^{\Lambda (r)}}{r}\, dr 
\end{equation} 

\noindent does provide the total gravitational mass of the star that enters the external 
(Reissner--Nordstr{\o}m) solution. The second term of this expression represents the electric 
coupling of charges in the star. It can be called the electromagnetic mass of the star 
(similar to the electromagnetic mass of an electron in the Lorentz theory). 

In the last expression, the second term  is an integral over the proper volume whereas 
the first is not. To correct this, we rewrite the expression for \( M\) in the form 

\begin{equation} 
M = m(R) = \int\limits_0^R {\rm e}^{-\Lambda (r)} \mu (r)\,   4\pi r^2 {\rm e}^{\Lambda (r)} dr + 
\int\limits_0^R \frac{q(r)\, \rho (r)}{r}\,  
4 \pi r^2 {\rm e}^{\Lambda (r)}\, dr .
\end{equation} 
Here, the integration in both terms is performed over the proper volume 
(\( 4\pi r^2 {\rm e}^{\Lambda (r)} dr\) being the element of the proper volume integrated 
over angle coordinates). The first term contains the factor 

\begin{equation} 
{\rm e}^{-\Lambda (r)} = \left[ 1 - \frac{2m(r)}{r} + \frac{q(r)^{2}}{r^{2}}\right]^{\frac{1}{2}} 
\end{equation} 

\noindent which represents the contribution of gravitational binding energy to the mass of the object.

The last expression for \( M\) clearly shows that gravity binds only the localized part of the 
mass (perfect fluid), but not the non--localizable part caused by electric coupling.  Any attempt to 
remove all mass except electromagnetic will produce an object that cannot be held together by gravity. 
The nature of the difficulties with the Lorentz model of an electron does not change in switching 
from Minkowski spacetime to the curved spacetime of general relativity.  

The only way to fix that would be to introduce some localized energy--momentum.  A common practice is to 
claim that vacuum should be thought of as a perfect fluid with an equation of state,

\begin{equation}
\rho + p = 0
\end{equation}

\noindent originally introduced by Zel'dovich \cite{Zel}. Such attempts, however, should be handled carefully.  
Zel'dovich used this equation (derived from the quantum electrodynamical consideration of zero--point energy) 
to argue for a non--zero value of the cosmological constant. His argument has some merits whether one agrees 
with it or not. However, arguments concerning the Lorentz model of an electron \cite{Gron} using the above 
equation of state (based on the quantum electrodynamic description of electrons and pairs \cite{Heitler}), cannot be made 
consistent. One may consider such objects for as long as they are not associated with actual electrons.  

It appears that general relativistic considerations in a static, spherically symmetric spacetime generated 
by a charged static, spherically symmetric object cannot mitigate difficulties with the Lorentz model of an 
electron, as the charge on the particles in the perfect fluid does not contribute to gravitational binding.  
Contrastingly, Wheeler's investigation of geons \cite{Geo,W} shows that it is possible to form pure 
electromagnetic objects held together by gravitational binding.  In the case of geons, gravity binds together 
 an electromagnetic field that is localizable.  This localizable EM field is described by the 
energy--momentum tensor and is not related to the electric coupling of charges.  According to Wheeler's work, 
such geons cannot be used to model electrons, at least classically, because of restrictions 
imposed on the mass of a geon.

Perhaps even more radical ideas are needed, such as Wheeler's ``charge without charge'' \cite{Geo,MW} that 
involve multiconnected spaces. Even then, such an attempt cannot be trivial in view of well--known difficulties 
with Wheeler's suggestions.

\appendix
\section{}
\setcounter{section}{1}

Here, we verify (\ref{EField}).  Let $ F^{0r} $ represent the 
radial electric field $ E^{r} $ as measured in the coordinate frame.  To compute the electric 
field inside of the star, we need the divergence of the electromagnetic field tensor

\begin{equation}
{F^{\mu\nu}}_{;\nu} = 4\pi J^{\mu}.
\end{equation}

\noindent Then,

\begin{eqnarray}
& & {E^{r}}_{;r} = {F^{0r}}_{;r} = 4\pi J^{0} \nonumber\\
&\textrm{where}& \:\: {F^{0r}}_{;r} = \frac{1}{\sqrt{-g}}[\sqrt{-g}F^{0r}]_{,r}.
\end{eqnarray}

Since $ J^{0} $ is measured in the coordinate frame, we write that $ J^{0} = \rho(r)u^{0} = \rho(r)e^{-\Phi(r)} $ 
\cite{TRK}.  This implies

\begin{equation}
\frac{1}{\sqrt{-g}}[\sqrt{-g}F^{0r}]_{,r} = 4\pi\rho(r)e^{-\Phi(r)}.
\end{equation}

In the coordinate frame, $ \sqrt{-g} = e^{\Phi + \Lambda}r^{2}\sin\theta $.  The left-hand 
side of (\ref{divE}) becomes

\begin{eqnarray}
& & \frac{1}{\sqrt{-g}}[\sqrt{-g}F^{0r}]_{,r} = \frac{1}{e^{\Phi + \Lambda}r^{2}\sin\theta}[e^{\Phi + 
\Lambda}r^{2}\sin\theta F^{0r}]_{,r} \nonumber\\
&=& \frac{1}{e^{\Phi + \Lambda}r^{2}\sin\theta}[e^{\Phi + \Lambda}r^{2}\sin\theta 
F^{\hat{0}\hat{r}}e^{-(\Phi + \Lambda)}]_{,r} \nonumber\\
&=& \frac{1}{e^{\Phi + \Lambda}r^{2}\sin\theta}[2r\sin\theta F^{\hat{0}\hat{r}} + 
r^{2}\sin\theta {F^{\hat{0}\hat{r}}}_{,r}].
\end{eqnarray}

\noindent Then,

\begin{eqnarray}
& & {F^{\hat{0}\hat{r}}}_{;r} = 2r\sin\theta F^{\hat{0}\hat{r}} + r^{2}\sin\theta {F^{\hat{0}\hat{r}}}_{,r} 
= 4\pi\rho e^{-\Phi}e^{\Phi + \Lambda}r^{2}\sin\theta \nonumber\\
&\Rightarrow& 2rF^{\hat{0}\hat{r}} + r^{2}{F^{\hat{0}\hat{r}}}_{,r} = 4\pi\rho e^{\Lambda}r^{2} \nonumber\\
&\Rightarrow& \frac{d}{dr}[r^{2}F^{\hat{0}\hat{r}}] = 4\pi\rho e^{\Lambda}r^{2}.
\end{eqnarray}

Here, we check that the derivative on the left-hand side is the same as the divergence as taken 
in the orthonormal frame.  The charge contained in a coordinate radius $ r $ is given by 
$ q(r) = \int_{0}^{r}4\pi\rho(r) r^{2}e^{\Lambda}dr $.  Integrating both sides with respect to 
the $ r $ coordinate,

\begin{equation}
r^{2}F^{\hat{0}\hat{r}} = \int_{0}^{r}4\pi\rho e^{\Lambda}r^{2}dr.
\end{equation}

We recognize the expression on the right as the charge contained in a coordinate radius $ r $; therefore,

\begin{eqnarray}
& & r^{2}F^{\hat{0}\hat{r}} = q(r) \nonumber\\
&\Rightarrow& F^{\hat{0}\hat{r}} = \frac{q(r)}{r^{2}}.
\end{eqnarray}

\appendix
\section*{Appendix B.}
\setcounter{section}{2}

This appendix was not in the original version of the paper. It was 
added in response to the questions of one of the referees of the original 
version of the paper who asserted that ``the analysis and the mass 
expressions (10), (22) depend on a coordinate system (2),'' 
which, in his opinion, made the final conclusions of the paper 
questionable. We wish to point out that similar arguments could be used 
to doubt the validity of the procedures involved in analysis of neutral 
spherically symmetric stars and of their total mass 
that can be found in standard texts on general relativity.
  
We discuss this issue below to show that our results and procedures are, 
in fact, coordinate independent and the ones used by everyone 
in the theory of neutral spherically symmetric stars. All the ingredients 
necessary for a proper understanding of this issue can be found in the 
literature \cite{MTW, Schutz}. We simply combined the ingredients and 
applied them to our paper. 

In our analysis, we use (as everybody else uses) 
a properly chosen slicing of spacetime by spacelike surfaces 
determined by comoving observers (sometimes called 
static observers) and the existence of spheres centered on the source of  
gravitational and electric fields in each slice. Such a slicing and spheres 
exist due to 
the symmetries of the problem (static object with spherical symmetry).   

A closer look at the mass expressions shows that both integrals in (22) are 
integrals of proper densities over the proper volumes of comoving observers. 
Presence of the Schwarzschild radial coordinate \( r\) in these expressions 
does not violate coordinate independence because (as is well known) this coordinate 
can be expressed in terms of the proper area \( A\) of an appropriate sphere 
\begin{equation} 
r = \sqrt{\frac{A}{4 \pi}}. 
\end{equation}
No matter what coordinates in the proper space (constant time slice) of the 
comoving observers are used, all the expressions for electric and 
gravitational fields in terms of the quantity \( r\), defined in 
this coordinate independent way, will remain the same as we have in our 
paper. This fact is used in deriving the expression for the electric 
field (7). 

Once again, nothing depends on coordinatization of proper spaces determined 
by comoving observers. Concerning the spacetime slicing by proper spaces 
of comoving observers (that happens to coincide with the surfaces of constant \( t\) 
in Schwarzschild coordinates), we would not recommend changing it to anything else as it 
is the appropriate slicing for defining the total mass (this is similar to 
choosing a comoving observer when defining the rest mass of a particle) and the 
only one in which the electric field remains static and purely electric. 
The reference to the 
Schwarzschild time coordinate  \( t\) should not be misinterpreted either 
because the only quantities present in all the expressions or derivations of 
these expressions are normalized vectors \(\partial /\partial t\) (i.~e.~the 
4-velocities of comoving observers). 

It is possible to write all expressions in terms of the 4-velocities 
of observers and of the proper areas of spheres.  We do not find this exercise useful, however. 
Instead, one can consider Schwarzschild coordinates \(r, t\) as shorthand 
notations for coordinate independent quantities, which is a common practice 
in the literature.
 
Identification of the quantity (22) with the total mass of a charged spherical star 
is based largely on the analysis of the orbits of test particles far from the 
source which shows that this quantity plays the part of mass in Kepler's 
third law. One can find a detailed description of this in standard texts on 
general relativity \cite{MTW}, and we do not think that we should pursue this issue further. 

To summarize, the quantity given by (22) does represent the total mass 
of a charged star and is not coordinate dependent.

\ack

MC would like to thank the NCSU Department of Mathematics Summer 2007 REG program 
for support of this work.

\end{document}